\documentclass[prl,aps,twocolumn,showpacs]{revtex4}
\usepackage{epsfig,amssymb,graphicx}

\begin{document}

\title
{Reply to the comment of Kolakowska and Novotny}
\author{Claudio Horowitz and Ezequiel V. Albano\\
Instituto de Investigaciones Fisicoqu\'{i}micas Te\'{o}ricas y Aplicadas (INIFTA), 
Universidad Nacional de La Plata, CONICET,
CCT-La Plata CONICET; Sucursal 4, CC 16 (1900) La Plata, Argentina}
\date{\today}

\begin{abstract}
In an early paper (Horowitz and Albano, Phys. Rev. E.,
{\bf 73} 031111 (2006)) we studied growing models, generically called $X/RD$,
such that a particle is attached to the aggregate with probability $p$ following the mechanisms of a 
generic model $X$, and at random [Random Deposition (RD)] with probability $(1 -p)$.
We also formulated scaling relationships that are expected to hold in the limits
$p \rightarrow 0$ and $L \rightarrow \infty $, where $L$ is the sample side.
In the previous comment, Kolakowska and Novotny (KN) stated that our scaling hypotheses does not hold. 
Here, we show that the criticisms of KN are outlined by analyzing data out of the proper scaling regime 
and consequently they are groundless and can be disregarded.
\end{abstract}
\maketitle

In the previous paper Kolakowska and Novotny (KN) \cite{KN} comment our
early paper on the "Dynamic properties in a family of competitive growing models"
\cite{our1}. Our paper, as well as a series of previous manuscripts on a related issue
\cite{we1,we2,we3}, addresses the properties of a wide variety of growing models, 
generically called $X/RD$, involving the deposition of particles
according to competitive processes, such that a particle is attached to the
aggregate with probability $p$ following the mechanisms of a generic model
$X$ that provides the correlations, and at random [Random Deposition (RD)]
with probability $(1 -p)$. A related study on that topic has also been published
by Braunstein et al. \cite{lidia}. 

The comments of KN are based essentially on two statements, both of them
erroneously attributed to us, which read as follows:

S1a) "The claim is made that at saturation the surface width $w(p)$ obeys a power-law 
scaling $w(p) \propto 1/p^{\delta}$, where $\delta$ is only either $\delta = 1$ 
or $\delta = 1/2$, which is illustrated by the models where $X$ is ballistic deposition 
and where $X$ is RD with surface relaxation." 
(Taken from the Abstract of the comment of KN \cite{KN}). A slightly different version of this 
statement can also be found in the comment of KN \cite{KN}, few lines below
equation (1), namely:

S1b) "The new claim that is being made in Ref.[1] (i.e. reference \cite{our1})
is that a nonuniversal and {\it model-dependent}  exponent $\delta$ in Eq.(1) must 
be only of two values, either $\delta = 1$ or $\delta = 1/2$, for 
models studied in Ref. [1]  (i.e. reference \cite{our1})."

On the other hand, the second statement, taken from the Abstract of the
comment of KN \cite{KN}, reads:

S2) "Another claim is that  in the limit $p \rightarrow 0$, for any lattice size $L$, 
the time evolution of $w(t)$ generally obeys the scaling 
$w(p) \propto (L^{\alpha} / p^{\delta}) F(p ^{2\delta}t/L^{z})$, where $F$ is 
Family-Vicsek universal scaling function."

Concerning both S1a) and S1b), our answer is that the statement 
is taken out of context from our paper, 
so it is incomplete and leads the reader to confusion. 
In fact, in our paper we explicitly state
in many places that the values $\delta = 1$ or $\delta = 1/2$ only "hold" in 
the {\bf $p \rightarrow 0$ limit}.
In fact, in our paper we state in the Abstract, the paragraph before the title of Section V,
in the title of Section V, and in the Conclusions that the universality in the $\delta$ and $y$ 
exponents should "hold" {\bf in the $p \rightarrow 0$ limit}.
Also, we have theoretically found those values of $\delta$ {\bf in the "$p \rightarrow 0$" limit} 
by using a correspondence between two neighboring sites in the discrete model $((h(i)-h(i+1)))$ 
and two types of random walks. 

Within this context it is worth mentioning that Braunstein et al. \cite{lidia}
have determined {\bf exactly} that $\delta = 1/2$ for a competitive model between 
Ballistic Deposition (BD) and RD, while  they found  $\delta = 1$ for
the competition between Random Deposition with Surface Relaxation (RDSR) and RD. 
Of course, Braunstein et al. \cite{lidia} clearly state that the
"scaling exponents derived are exact for  $p \rightarrow 0$." They also recall that 
"at finite $p$, we expect deviation from the exact scalings which indeed were observed
numerically (in reference \cite{we2}." See the last paragraph of Section II, page 2.

Figures 1(a) and 1(b) of the comment of KN \cite{KN} show the 
behavior of various models, with $X  \equiv  RDSR$, 
$X \equiv  BD$ and some variants. Results are shown for a wide 
range of $p$, actually most results correspond to $p \ge 0.3$, i.e., far away from the 
correct scaling regime given by the {\bf $p \rightarrow 0$ limit}. 
Subsequently, by describing the figures, they state that "in special cases an approximate 
power-law  $w(p) \propto 1/p^{\delta}$  may be observed, however, this is not a principle."
From our point of view, it is obvious that one would not expect a nice power-law fit of the 
data within the whole range of $p$, but {\bf "only in the limit of $p \rightarrow 0$."} 
Also, a careful inspection of figures 1(a) and 1(b) of the comment nicely shows that 
the power-law predicted in our paper fits very well the data for {\bf $p \rightarrow 0$!}.  
So, the data shown in figure 1 of the comment of KN \cite{KN} seem to be correct, 
but they are far away from the right scaling limit. The figure clearly suggests that for 
data taken correctly, namely, for $p << 0.1$ as stated in our paper \cite{our1}, 
they would certainly obey the proposed scaling behavior.

In a related context, also at the end of the subsection {\bf Saturation}, KN state that \cite{KN}
"The other two examples shown in fig. 1 defy a linear fit. In these cases there 
is no power law of the type claimed in Ref.[1]" (i.e., reference \cite{our1}). This 
absence of power-law scaling in $p$ is also evident in fig.4 of Ref.[1] (i.e. reference \cite{our1}).
We notice that in contrast to that opinion, an excellent power-law behavior can be observed 
in figure 4 of our paper \cite{our1}, but of course in the {\bf right scaling regime,
namely, for $p \rightarrow 0$}. 

Concerning S2, our answer is also that the statement is incomplete and
incorrectly formulated, so it leads the reader to confusion. 
In fact, in reference \cite{our1} we clearly state that S2 only 
holds, as is well known for the standard scaling relations, in the 
{\bf $L \rightarrow \infty$ limit}, but not "for any lattice size $L$," 
as in the misleading statement of KN. The right scaling regime for the 
sample size is stated in many places of our paper, e.g. along the
explanations of Section II, as well as the explicit scaling indications
written in equations (6), (7) and (8) \cite{our1}.   
We only mentioned in Section III that there is numerical evidence of negligible 
finite-size corrections to the value of the exponent $\delta$. Obviously, this property
of $\delta$ is not equivalent to the statement formulated by KN in a general way.

All discussed issues and the remaining topics of the comment suggest that KN 
did not understand the concept of scaling in the $p \rightarrow 0$ and $L \rightarrow \infty $ limits, 
and therefore they are unable to understand either the correct content of our paper 
or their own results. 
For example, in the subsection {\bf The RD limit},  another comment of KN reads:
"Another claim of Ref.[1] is that Eq.(1) with the power-law prefactors $p^{\delta}$
where ($\delta = 1$ or $1/2$) would prevail in the limit $p \rightarrow 0$, and 
that such a scaling would be universal. We tested these claims in simulations of 
RD+BD models and found the evidence to the contrary (fig.2-3)."
Again, figures 2 and 3 of the comment of KN show the behavior of 
RD+BD models using $p$ in the range $[0.1 ,1]$ !, which 
of course is not at all close to the limit $p \rightarrow 0$. 

We notice that it is easy to understand why KN claim that our scaling did not fit 
the data very well. This is the obvious result that 
one may always obtain just by working with data taken out of the scaling regime but 
using values for the exponents corresponding to the correct regime. In fact, as we 
have already shown in our paper, as well as in previous work \cite{we1,we2}, 
when using small samples, data collapse can only be obtained by using the effective 
values of the exponents according to the range of $p$ and $L$ used.

In summary, our paper shows that for some models and 
within the $p \rightarrow 0$ and  $L \rightarrow \infty$  limits, the proposed 
scaling is a universal principle. On the other hand, in their comment, 
KN have shown that data taken out of the correct scaling limit may depart from scaling.

We acknowledge financial support from the Argentinian Science
Agencies CONICET and ANPCyT, and the UNLP.


\end{document}